\documentclass{article}

\usepackage{microtype}
\usepackage{graphicx}
\usepackage{subfigure}
\usepackage{booktabs} % for professional tables

\usepackage{hyperref}

\usepackage[accepted]{icml2025}

\usepackage{amsmath}
\usepackage{amssymb}
\usepackage{mathtools}
\usepackage{amsthm}

\usepackage[capitalize,noabbrev]{cleveref}

\theoremstyle{plain}

\theoremstyle{definition}

\theoremstyle{remark}

\usepackage[textsize=tiny]{todonotes}

\icmltitlerunning{\sys: High-Throughput MoE Inference on a Single GPU with Module-Based Batching}

\usepackage{float}
\usepackage{comment}
\usepackage{xspace}
\usepackage{multirow}
\usepackage{subcaption}
\usepackage[skip=8pt]{caption}
\usepackage[nobiblatex]{xurl}

\newcommand{\sys}{\textsc{MoE-Gen}\xspace}

\newcommand{\tinyskip}{}
\newcommand{\mypar}[1]{\tinyskip\noindent\textbf{#1.}\xspace}

\def\-{\raisebox{.75pt}{-}}
\DeclareMathSymbol{\shortminus}{\mathbin}{AMSa}{"39}

\usepackage{mathtools}
\newcommand{\eg}{\text{e.g.,}\ }

\newcommand{\ie}{\text{i.e.,}\ }

\newcommand{\fixme}[1]{\textbf{\color{red}[#1]}}

\ifdefined\RELEASE
  \newcommand{\ignore}[1]{}
  \newcommand{\fixme}[1]{}
  \newcommand{\yao}[1]{}
  \newcommand{\luo}[1]{}
  \newcommand{\leyang}[1]{}
  \newcommand{\zhan}[1]{}
  \newcommand{\TODO}[1]{}
  
\else
  \newcommand{\ignore}[1]{}
  \newcommand{\yao}[1]{{\textcolor{orange}{[~YAO:~#1~]}}}
  \newcommand{\luo}[1]{{\textcolor{gray}{[~LUO:~#1~]}}}
  \newcommand{\leyang}[1]{{\noindent\textcolor{blue}{[LX:~#1]}}}
  \newcommand{\zhan}[1]{{\textcolor{cyan}{[~Zhan:~#1~]}}}
  \newcommand{\TODO}[1]{{\textcolor{red}{TODO:~#1}}}
  
\fi

\newcommand{\algorithmstyle}[1]{\renewcommand{\algocf@style}{#1}}

\newlength\myindent
\usepackage{chngcntr}
\usepackage{ulem}

\begin{document}

\twocolumn[
\icmltitle{\sys: High-Throughput MoE Inference on a Single GPU \\ with Module-Based Batching}



\icmlsetsymbol{equal}{*}

\begin{icmlauthorlist}
\icmlauthor{Tairan Xu}{equal,edi}
\icmlauthor{Leyang Xue}{equal,edi}
\icmlauthor{Zhan Lu}{edi}
\icmlauthor{Adrian Jackson}{EPCC}
\icmlauthor{Luo Mai}{edi}

% \icmlauthor{Firstname5 Lastname5}{edi}
\end{icmlauthorlist}

\icmlaffiliation{edi}{The University of Edinburgh}
\icmlaffiliation{EPCC}{EPCC, The University of Edinburgh}

\icmlcorrespondingauthor{Tairan Xu}{t.xu-29@sms.ed.ac.uk}
\icmlcorrespondingauthor{Leyang Xue}{leyang.xue@ed.ac.uk}
\icmlcorrespondingauthor{Zhan Lu}{z.lu-64@sms.ed.ac.uk}
\icmlcorrespondingauthor{Adrian Jackson}{a.jackson@epcc.ed.ac.uk}
\icmlcorrespondingauthor{Luo Mai}{luo.mai@ed.ac.uk}


\icmlkeywords{Machine Learning Systems, ICML}

\vskip 0.3in
]



\printAffiliationsAndNotice{\icmlEqualContribution} % otherwise use the standard text.

\begin{abstract}
This paper presents \sys, a high-throughput MoE inference system optimized for single-GPU execution. Existing inference systems rely on model-based or continuous batching strategies, originally designed for interactive inference, which result in excessively small batches for MoE’s key modules—attention and expert modules—leading to poor throughput. To address this, we introduce module-based batching, which accumulates tokens in host memory and dynamically launches large batches on GPUs to maximize utilization. Additionally, we optimize the choice of batch sizes for each module in an MoE to fully overlap GPU computation and communication, maximizing throughput. Evaluation demonstrates that \sys achieves 8–31× higher throughput compared to state-of-the-art systems employing model-based batching (FlexGen, MoE-Lightning, DeepSpeed), and offers even greater throughput improvements over continuous batching systems (e.g., vLLM and Ollama) on popular MoE models (DeepSeek and Mixtral) across offline inference tasks. MoE-Gen’s source code is publicly available at https://github.com/EfficientMoE/MoE-Gen
\end{abstract}

\section{Introduction}
MoE architectures are increasingly favoured in LLMs because their router-based design activates only a subset of experts per token, reducing computational overhead and making them more suitable for deployment on personal machines with limited GPU resources. For locally deployed MoEs, AI developers often run high-throughput inference tasks such as benchmarking\cite{chatbot-arena,gsm8k} to evaluate fine-tuned models for personal AI applications, data wrangling\cite{DBLP:journals/pvldb/NarayanCOR22,DBLP:journals/pvldb/RenenSK24} for cleaning datasets and extracting information, and feature extraction~\cite{mischler2024contextual,DBLP:conf/acm/AsaiMZC23} to generate embeddings for retrieval-based LLMs and other downstream applications. Unlike conventional interactive inference tasks (e.g., ChatBot), high-throughput inference trades off lower latency for higher batch sizes, optimizing overall throughput.

A major challenge for high-throughput MoE inference is the model’s large size, which often exceeds a single GPU’s memory capacity. To overcome this, AI developers use memory offloading, where the full model parameters of an MoE and its KV-cache are stored in host memory—typically much larger and more cost-effective to expand than GPU memory. Parameters and KV-cache are then fetched into the GPU only when activated.

\begin{table}[t]
    \centering
     \resizebox{\linewidth}{!}{%
    \begin{tabular}{|c|ccc|ccc|}
    \hline
      & \multicolumn{3}{c|}{Prefill Expert Avg.} & \multicolumn{3}{c|}{Decoding Expert Avg.} \\
    \cline{2-7}
    & Bsz        & Util  &TP    & Bsz       & Util       &TP     \\
    \cline{2-7}
    \hline

        DeepSpeed               &153        &52\%       &109       &0.3         &0.1\%                       &1    \\
    FlexGen*                &115        &49\%           &77          &0.3       &0.1\%                     &1     \\
        MoE-Lightning*          &134        &50\%           &98          &0.4       &0.1\%                     &1   \\
    \hline

        \textbf{\sys} &\textbf{8192}    &\textbf{100\% }  &\textbf{841}  &\textbf{75}  &\textbf{41\% }   &\textbf{31}     \\
    \hline
    \end{tabular}
    }
    \caption{Offloading throughput (TP in tokens/s) is measured for \textit{DeepSeek-V2 236B} on an NVIDIA A5000 (24GB) with 512GB of host memory and a context length of 768 tokens (512 for the prompt, 256 for decoding). \sys's module-based batching enables up to a 2$\times$ increase in GPU FLOPs utilization and a 7.7-11$\times$ improvement in throughput. During the decoding phase, \sys achieves a 31$\times$ improvement in throughput. `Bsz' denotes the average number of tokens routed to an expert.}
    \label{tab:intro-example}
\end{table}

When offloading is enabled, high-throughput LLM inference systems often suffer from low GPU utilization, leading to suboptimal throughput performance. These systems typically use model-based batching, where the entire MoE model processes a batch of input tokens at the model ingress, but within the model, each expert handles only a small fraction of tokens assigned by the router. This results in extremely low GPU utilization during decoding. We illustrate this in~\cref{tab:intro-example}.
Systems like FlexGen~\cite{flexgen}, DeepSpeed-Inference~\cite{ds-infer}, Mixtral-Offloading~\cite{mixtral-offload}, and MoE-Lightning~\cite{moe-lightning} often operate with batch sizes 40–1000× smaller than what is needed to fully utilize a GPU during LLM decode, significantly reducing throughput compared to the prefill phase. While continuous batching~\cite{orca}, as used in interactive inference systems like Llama.cpp~\cite{ollama} and vLLM~\cite{vllm}, could improve GPU utilization, our analysis shows that it is optimized for time-to-first-token (TTFT) rather than throughput. In practice, it leads to even smaller batch sizes during decoding, further lowering throughput. 

In this paper, we explore methods to significantly improve high-throughput MoE inference on a single GPU. Our key idea is that MoE models have only two compute-intensive modules: attention and experts. For those two modules, we can accumulate sufficient tokens in the CPU’s host memory to form large batches and ensure that GPUs process them only when the batch size is large enough to fully utilize GPU resources, increasing throughput. Additionally, we carefully optimize batch size so that GPU computation and memory communication are fully overlapped, keeping GPU utilization high.
This approach is feasible because CPU memory is significantly cheaper than GPU memory and is typically large enough to store the entire MoE model in an offloading scenario. Moreover, by processing larger batches per module, we reduce the need for repeated host-to-GPU transfers, alleviating I/O bottlenecks. We call this new design \emph{module-based batching}.

Building on this idea, we design \sys, a high-throughput MoE inference system that fully utilizes a single GPU. Our key contributions include:

\mypar{Contribution 1} We propose a module-based batching strategy for MoE architectures, where attention and expert modules are carefully selected and organized to incrementally build large batches, maximizing GPU utilization.

 \mypar{Contribution 2} We propose an optimized system design for high-throughput MoE inference on a single GPU. This design includes full support for module-based batching, enhanced optimizations for managing offloaded KV-cache and model weights, and parallel CPU cores to offload partial computations from the GPU, further improving throughput.
 
 \mypar{Contribution 3} We formulate an optimization problem to determine the optimal batch size for different attention and expert modules, considering various practical factors such as MoE model architecture (e.g., expert size and count), hardware capabilities (e.g., GPU memory size), and system parameters (e.g., buffer size and peak memory consumption). To solve this, we propose a search policy that efficiently finds the optimized batch size for both prefill and decode phases.

We evaluated \sys against extensive baseline systems including FlexGen, DeepSpeed-Inference, MoE-Lightening, vLLM and Ollama (llama.cpp) using popular, open-sourced MoE models, including DeepSeek-V2~\cite{deepseek-v2} and Mixtral~\cite{mixtral} with benchmarks such as ChatBot-Arena~\cite{chatbot-arena}, LongBench~\cite{longbench}, MMLU~\cite{mmlu} and GSM8K~\cite{gsm8k}. In the evaluation, \sys achieves 9-63$\times$ less time to complete the inference on datasets with 8K–116K prompts, 16-33$\times$ higher decoding throughput across different models on a single commodity GPU compared to model-based offloading systems, and 7.7-11$\times$ higher prefill throughput on MoE models with higher sparsity (e.g., DeepSeek). Additionally, \sys delivers 7-13$\times$ throughput improvement for long-context generation (6K–24K context length), fully leveraging the capabilities of state-of-the-art MoE LLMs.

\section{Background}
\label{sec:background}
\begin{figure}[t]
    \centering
    \includegraphics[width=\linewidth]{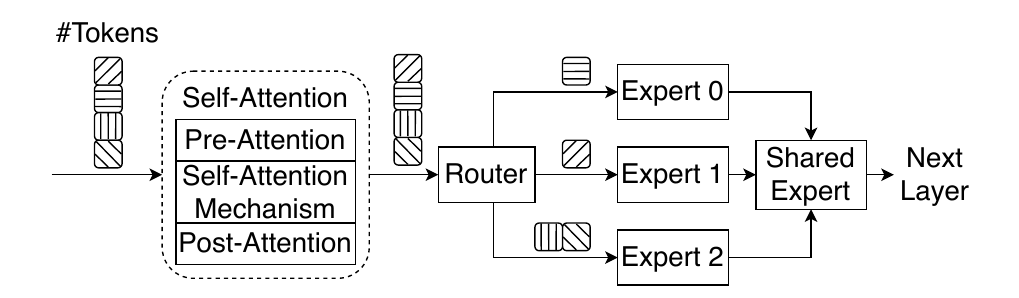}
    \vspace{-0.25in}
    \caption{Illustration of one layer in MoE models.}
    \label{fig:sec-2}
\end{figure}

\mypar{MoE inference}
We describe the inference process of the MoE model as shown in~\cref{fig:sec-2}.
A typical layer in MoE models consists of a self-attention layer followed by a sparse MoE layer. The input tokens to a layer are first processed by the self-attention layer, which can be generally divided into three stages: the pre-attention stage (\eg QKV projection), the self-attention mechanism stage (\eg $QK^{T}$), and the post-attention stage (\eg output projection). 
After the self-attention layer, tokens are passed to the sparse MoE layer, where a router assigns each token to a subset of experts, typically using a top-$k$ selection strategy. 
Each token is processed by $k$ selected experts, and the final output is obtained by computing a weighted average of the outputs from these experts. 
Some model architectures, such as DeepSeek-V2 and Qwen2MoE \cite{qwen-moe}, incorporate a shared expert that all tokens pass through. 
The processed tokens then proceed to subsequent layers in the model. 
We omit layer normalization and residual connections in our discussion, as their exclusion does not affect the key structure.

MoE batched inference follows the same procedure as LLM generative inference, which operates in two phases:
i) prefill: A batch of prompts is processed to generate the KV-cache at each attention layer.
ii) decoding: New tokens are generated in an auto-regressive manner. The output tokens from the previous forward pass are used as the input tokens for generating the next token. In each forward pass, the KV-cache for the new input token is generated and appended to the existing KV-cache, forming the complete context so far. The computational intensity in the decoding phase is typically orders of magnitude lower than in the prefilling phase, as only one token per sequence is passed into the model.

\mypar{MoE offloading}
The offloading system typically manages two levels of memory: CPU memory for excess model weights and key-value (KV) states, and GPU memory for computation and fast data access. When model weights are required for GPU computation, they can either be fetched in advance (overlapped with other computations) or fetched on-demand. A resident store can be designed in GPU memory to persistently hold model weights and/or KV-cache, while a staging \textit{buffer} is used to prefetch dynamic data. If the GPU attempts to compute with data (e.g., weights or KV states) that are not yet in its resident store, it must stall until the data are transferred from CPU memory.

In offloading systems, the bandwidth between the host and memory is often a scarce resource. Leveraging CPU computation resources to process data locally can potentially increase overall throughput~\cite{moe-lightning}.

\section{Related Work}
\begin{figure}
    \centering
    \includegraphics[width=\linewidth]{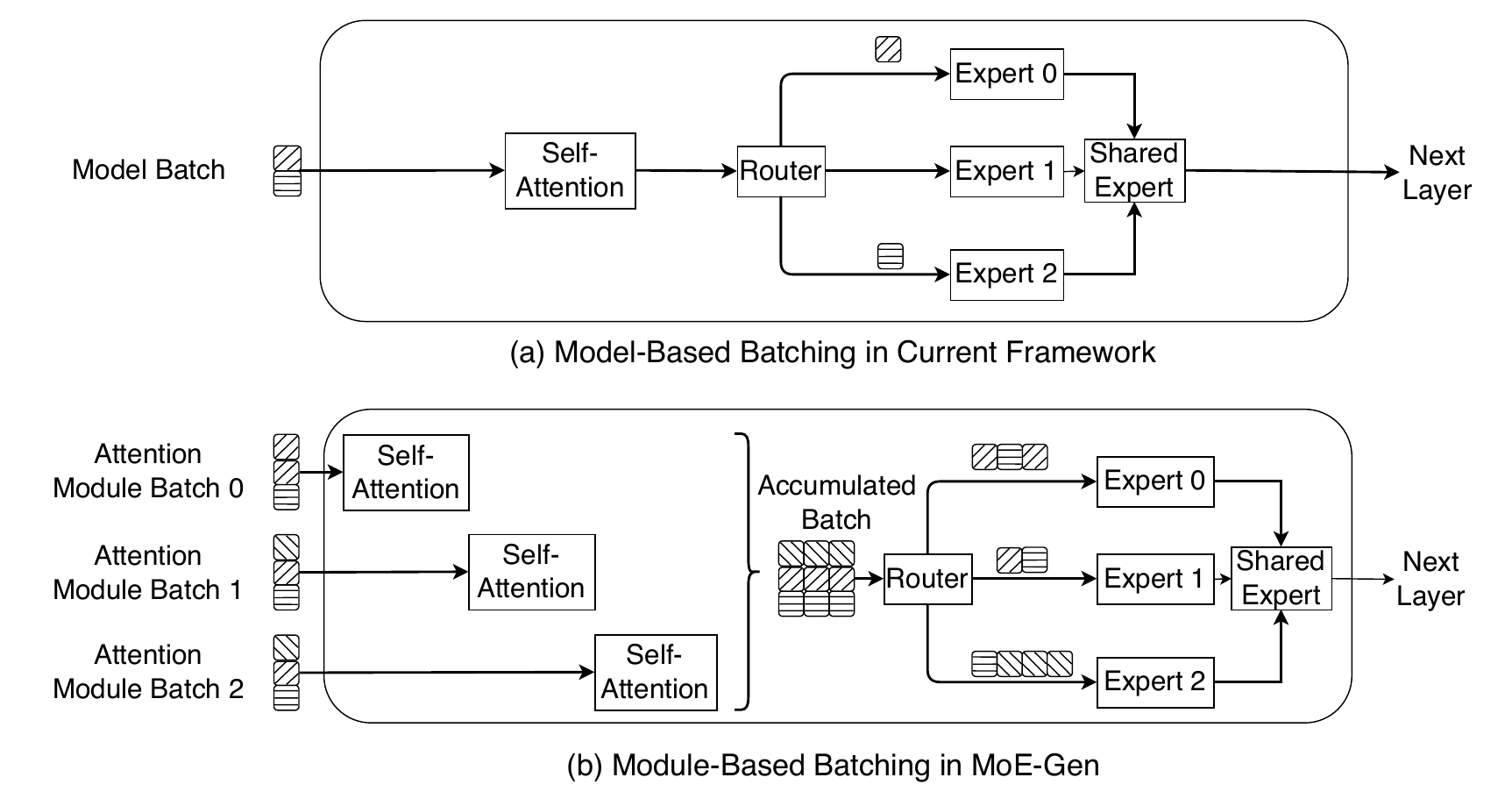}
    \vspace{-0.25in}
    \caption{Model-based batching employs a single, unified batch size throughout the entire forward pass, whereas module-based batching iteratively process modules with small batches to form larger batches.}
    \label{fig:related-work}
\end{figure}

We show the commonly applied model-based batching and \sys's module-based batching in~\cref{fig:related-work}.
\mypar{(1) Model-based batching}
DeepSpeed-Inference~\cite{ds-infer} and FlexGen~\cite{flexgen} are designed for dense transformer models and treat MoE layers as dense MLP layers, resulting in insufficient batch sizes for expert layers. FlexGen processes multiple rounds of forward passes reusing the same fetched model weights. In each forward pass, a unified batch is propagated through the entire model, without addressing the batch size limitation for MoE experts. MoE-Lightning~\cite{moe-lightning} improves performance over FlexGen by optimizing GPU-CPU-I/O overlap but retains the same batching strategy.
Mixtral Offloading~\cite{mixtral-offload} supports the offloading of Mixtral-series of MoE models, making them popular among users with limited GPU resources. 

\mypar{(2) Continuous batching for high throughput} Continuous batching is orthogonal to both model-based and module-based batching, as it operates at the sequence level. Each forward pass still relies on model-based batching.

Continuous batching frameworks often insert small prefill batches (frequently of size 1) into the decoding phase, leading to an even smaller average batch size over the entire execution.

Frameworks supporting continuous batching include vLLM~\cite{vllm}, TensorRT-LLM~\cite{trt-llm}, and Llama.cpp~\cite{ollama}. NEO~\cite{neo} interleaves prefill and decoding across the GPU and CPU, while systems such as BlendServe~\cite{blend-serve} and others~\cite{stream-batch} share the GPU in the temporal domain using micro-batches—ultimately facing the same issue as vLLM. In offloading scenarios, continuous batching performs even worse than traditional model-based batching. Therefore, we exclude it from further discussion in this paper and only report the result for reference.

\mypar{(3) Batching in training systems}
Training systems often interoperate fixed global batch sizes to ensure accuracy.
They seek to reduce communication overhead between GPUs over bottlenecked links~\cite{janus,lina,smart-moe}.
This is orthogonal to inference systems where batch size can vary without affecting the model quality.
Training systems only have prefilling~\cite{alpa}, while decoding is not the major concern.

\mypar{(4) Interactive inference systems with offloading}
Pregate-MoE~\cite{pregate-moe}, ExpertFlow~\cite{expert-flow}, MoE-Infinity~\cite{moe-infinity}, ProMoE~\cite{pro-moe} and BrainStorm~\cite{brainstorm} use predictors to instruct expert prefetching before batched inference.
While experts are often densely activated under large batch sizes, the prediction-based optimization becomes unnecessary for throughput optimizations.
Fiddler~\cite{fiddler} and PowerInfer~\cite{powerinfer} support running attention or experts on CPU to alleviate the I/O bottleneck, while mainly optimized for latency in small batch sizes.
% \mypar{(4) Expert quantization}
EdgeMoE~\cite{edge-moe}, AdapMoE~\cite{adap-moe}, and Hobbit~\cite{hobbit} reduce I/O traffic by replacing high-precision expert parameters with quantized versions. 
These methods may incur accuracy loss as sparsity increases~\cite{sparse-quant}.

\section{Module-Based Batching}
\subsection{Design Intuition}

We aim to design a batching engine and strategy that accomplish two key objectives:
(i) ensuring sufficient large input batches for modules to fully utilize GPU FLOPs, and  
(ii) overlapping computation and fetching times to reduce the end-to-end execution time.

\begin{figure}[t]
    \centering
    \includegraphics[width=\linewidth]{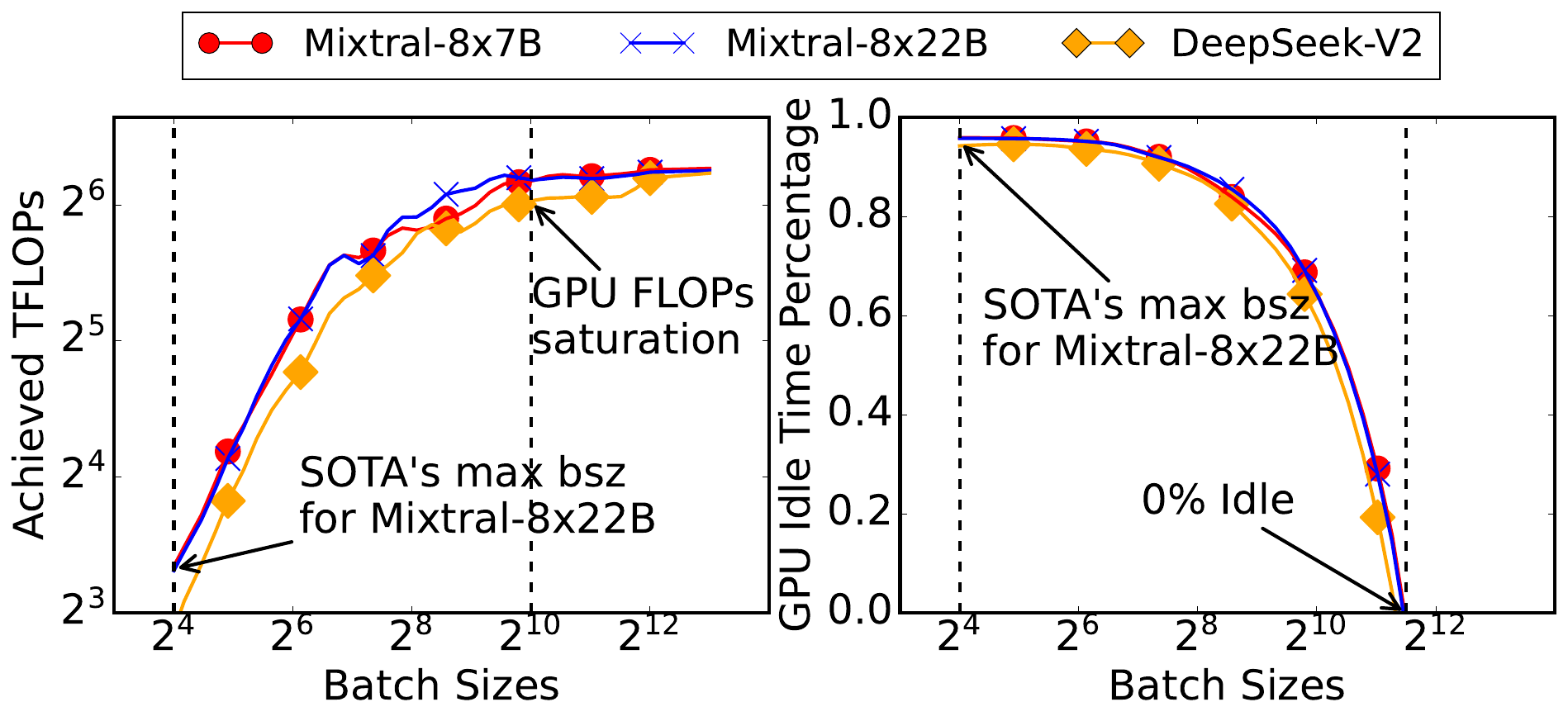}
    \vspace{-0.25in}
    \caption{Left: Achieved FLOPs in the non-offloading scenario. This metric represents the number of floating-point operations performed by an expert module, normalized by the GPU compute time.
    Right: Percentage of GPU idle time in the offloading scenario on an NVIDIA A5000 (PCIe 4.0, 32 GB/s). This metric measures the ratio of the expert module’s execution time to the time required to transfer the necessary weights from the CPU to the GPU.
    }
    \label{fig:expert-arithmetic-intensity}
\end{figure}

\mypar{Necessity of large batch size}
We make a key observation: the batch size for model-based batching is constrained by the module with the highest memory usage—often the attention module. However, because each expert has low arithmetic intensity, a substantially larger batch size is required to achieve optimal GPU utilization.

As shown in~\cref{fig:expert-arithmetic-intensity} (Left), at least $2^{10}$ tokens are required to fully utilize GPU compute, whereas the average number of input tokens per expert in current SOTA models is only $2^{4}$.  
Furthermore, in~\cref{fig:expert-arithmetic-intensity} (Right), we illustrate a common scenario in offloading: the computation of an expert should be fully overlapped with the fetching of the next expert, resulting in zero GPU idle time.
In this case, more than $2^{11}$ input tokens per expert are needed to ensure that the GPU does not remain idle while waiting for memory transfers.  
Thus, aggregating a larger batch size for experts is essential to achieve optimal performance.
Similar phenomenon has also been observed in the forward pass of attention module in the decoding phase.

\mypar{Necessity of searching batching strategy}
While our main goal is to improve device utilization by increasing input batch sizes for modules, other factors can also affect the execution time of a forward pass. Their influence on throughput is primarily indirect, operating through the utilization or conservation of resources (\eg GPU memory, PCIe bandwidth). For example, applying CPU-based computation to tasks that remain in host memory can save scarce HtoD memory bandwidth. However, whether this saving translates into throughput gains depends on (1) the CPU's computation speed and (2) whether we have reserved sufficient GPU buffer to effectively leverage the saved bandwidth. 
Similarly, if a module requires complex memory reshaping or migration, it may introduce delays but can also provide opportunities to overlap memory copy operations. Appropriate configurations must be chosen to seize this opportunity. 
Consequently, a searching strategy is necessary to determine the resulting throughput and select the optimal configurations.

\mypar{Means to facilitate large batch size}
We need to ensure that the GPU memory is sufficient to process large amounts of input in a single module. During module execution, design-dependent intermediate states (e.g., QKV projections in standard attention or the up-projection result of compressed KV-cache in DeepSeek) can consume significant GPU memory at runtime, which constrains the achievable batch size. To address this, \sys aims to keep fewer model weights and less KV-cache data on the GPU, thereby facilitating the use of larger batch sizes.

\subsection{\sys Engine Design}
\mypar{Module-based batching}
We propose the batching strategy of \sys as shown in~\cref{fig:related-work}.
As we observe different modules have distinct memory and FLOP requirements, 
\sys instead assigns different batch sizes to each module, \ie attention and expert.
We choose these two components as the base for batching, since the attention module present higher memory demand, which is suitable for lower batch size.
Conversely, the expert needs to scale to larger batch size, presenting two extremes.
\sys then accumulates multiple attention batches and processes them in one at the expert module stage, effectively increasing the batch size for the expert module.

\mypar{Sequential execution of experts}
Under large batch sizes, the number of tokens routed to each expert is often uniformly distributed, as observed in previous work and by design of the MoE auxiliary loss~\cite{moe-infinity,mixtral}. Consequently, \sys does not rely on heuristics or prediction-based prefetching. Instead, it focuses on enabling large batch sizes and prefetching, executing experts in a sparse MoE layer sequentially.

\begin{figure}[t]
    \centering
    \includegraphics[width=\linewidth]{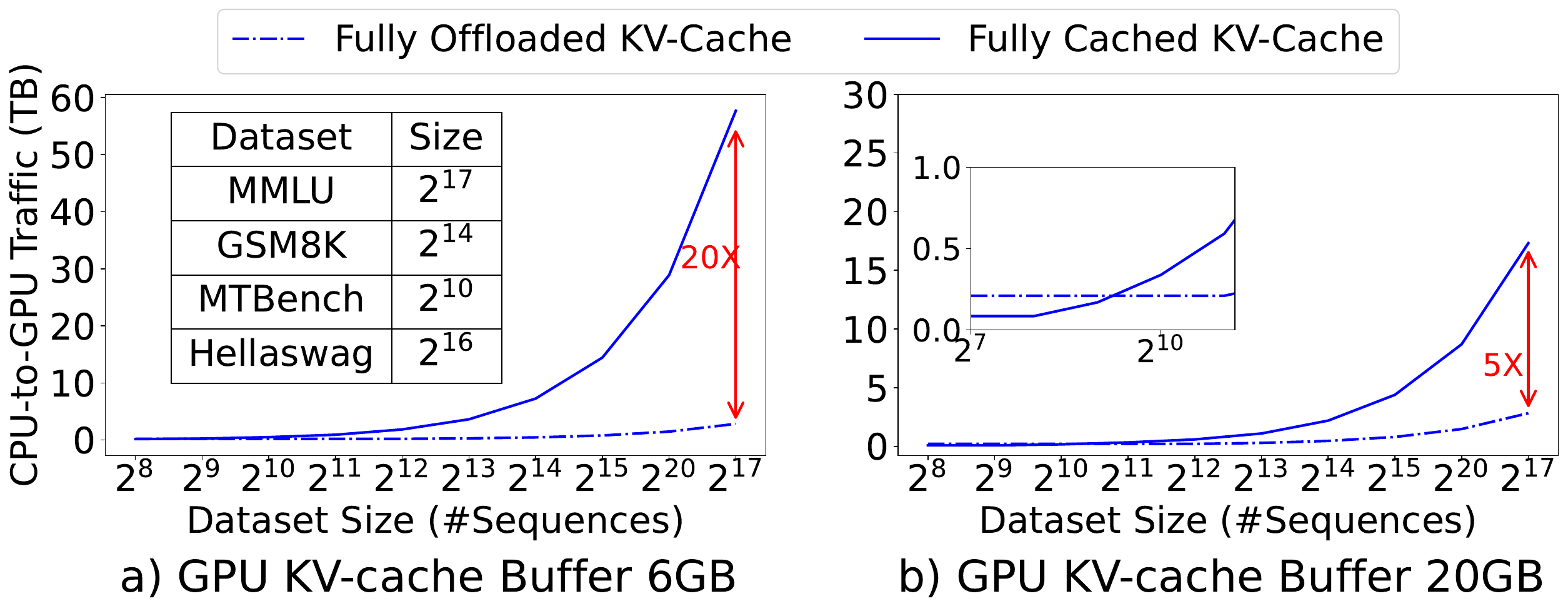}
    \vspace{-0.25in}
    \caption{Fetching traffic over dataset, showing fully offload KV-cache benefits performance. Using Mixtral-8x7B with CPU KV-cache capacity 128GB. We pad/truncate each prompt to same length and decode same length.
    }
    \label{fig:offload-traffic}
\end{figure}

\mypar{Full KV-cache offloading}
We initially consider KV-cache partial offloading~\cite{vllm}. 
However, we demonstrate that fully offloading the KV-cache outperforms partial offloading, particularly when considering the completion time for processing an entire dataset. 
The primary reason is that caching the KV-cache in GPU memory limits batch size, leading to increased fetching traffic for expert weights (e.g., up to 86GB for Mixtral-8x7B). By trading off KV-cache copying, \sys achieves up to 20× savings in fetching traffic, as shown in~\cref{fig:offload-traffic}. 
While smaller datasets may benefit from retaining KV-cache in GPU memory, especially as GPU memory capacity increases (shown in~\cref{fig:offload-traffic} (b)), popular benchmarking datasets are typically orders of magnitude larger, making KV-cache offloading more advantageous.

\mypar{Single GPU buffer for dense modules}
In MoE models, in contrast to sparse activated experts, there are dense modules that would be activated for each token (\eg attention modules and shared experts in DeepSeek). We find that setting the GPU prefetch buffer size to the size of dense modules in a single layer is sufficient to create overlapping.
The fetching of such modules only has two cases: 
(i)~\textit{with sufficient HtoD bandwidth}: dense modules can be prefetched and fully overlapped;
(ii)~\textit{without sufficient HtoD bandwidth}: as bandwidth is fully occupied by expert fetching, dense modules need to be fetched on demand.
In both cases, the buffer can be cleared and repurposed to the next dense module. Empirically, we verified that assigning more buffer space to dense modules would not increase throughput. When they are large, they could downgrade performance by squeezing the space for other components.

\mypar{CPU for self-attention} 
Matrix multiplications on the CPU (\eg attention projection and expert) are often 10-100x slower than computation on the GPU even accounting for the fetching time of weights~\cite{gemm-nvidia,fiddler}.
Due to the arithmetic intensity in matrix-vector multiplications (GEMV) in $QK^T$ operation, CPU can process data at a pace comparable to the time required to transfer data with PCIe4.0 to the GPU and perform computations there.~\cite{powerinfer,moe-lightning}.
We only use the CPU to compute self-attention mechanism in \sys.
This requires a custom CPU kernel to be implemented with better cache performance than current PyTorch and CBLAS, similar to the CPU version of FlashAttention~\cite{flashattn-2}.

\begin{figure}[t]
    \centering
    \includegraphics[width=.8\linewidth]{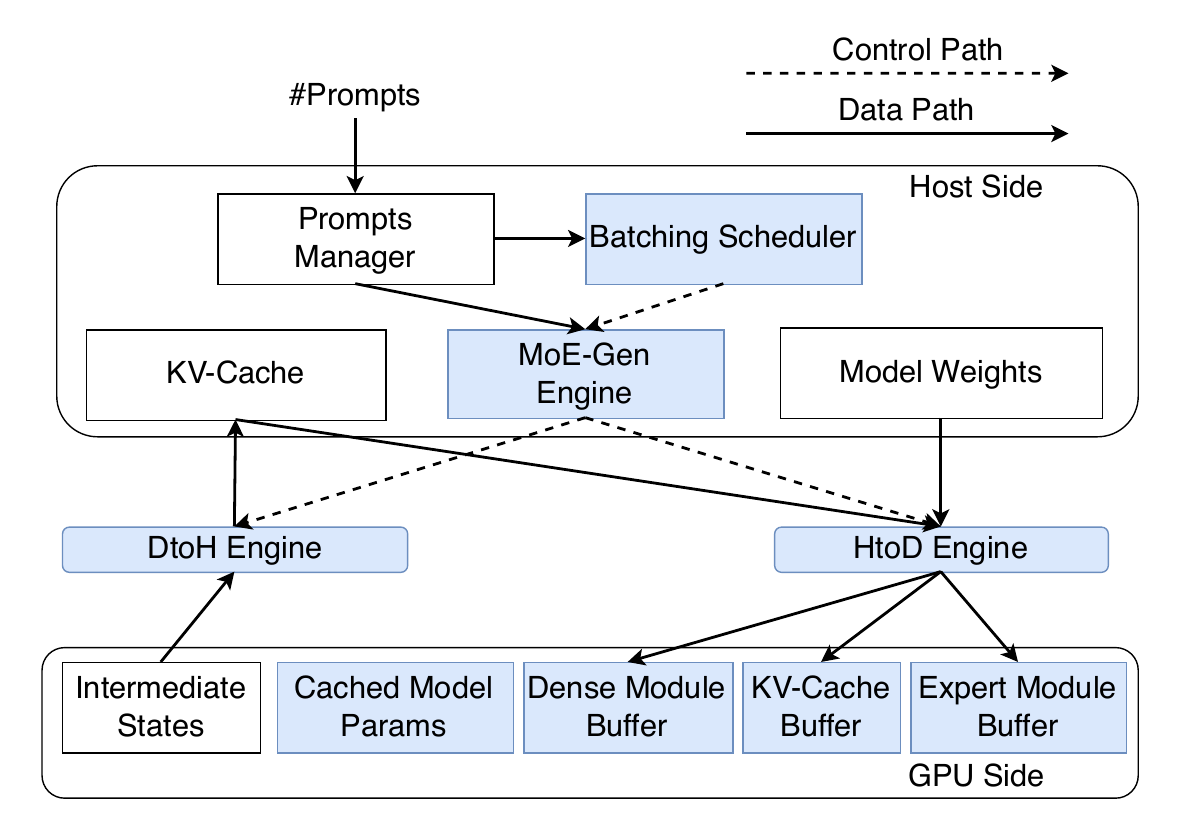}
    \vspace{-0.15in}
    \caption{\sys system components.}
    \label{fig:system-component}
\end{figure}

\mypar{System components}
Our design choices has led to the system architecture shown in~\cref{fig:system-component}.
\sys features a batching scheduler, creating batching strategy based on hardware (\eg connection speed, GPU memory capacity) and software (\eg performance and memory usage of GPU and CPU kernels under various input batch sizes) profiling. 
Using this information, the scheduler enumerates candidate configurations in the search space and applies them to the DAG constructor to estimate the overall runtime of each configuration. 
It then selects the configuration with the shortest completion time and sends its decision to the \sys Engine, which is responsible for executing the inference. At this point, the KV-cache buffer, expert module buffer, and dense module buffer are allocated in GPU memory based on the selected configuration’s size requirements.

The \sys Engine launches batched module execution and submits batched memory copy tasks to the HtoD engine in advance. 
The engine accumulates batches at each attention layer and MoE layer.
Meanwhile, KV-cache offloading and update tasks are submitted to the DtoH engine at runtime. The \sys Engine also manages all necessary synchronization between computation and memory-copy operations. Details about workload profiling and implementation of the engine are in \cref{sec:appendix-implementation}.

\subsection{Batching Strategy Formulation}
\mypar{Problem Formulation}
The batching strategy is working under the following hardware settings.
We consider a machine with two devices: a GPU and a CPU, both of which have memory capacity and computational capabilities.
The two levels of memory hierarchy are interconnected with two links HtoD link and DtoH link, which only does data transmission.
As the MoE model and corresponding KV-cache cannot fit entirely within the GPU, we offload them to Host memory that is close to the CPU.

Our aim is to find the module-based batching strategy that leads to the maximum throughput under the given system capability.
Equivalently, we look for an accumulated batch size $B$ constrained by the memory capacity and corresponding minimal runtime $T$ of the batch.
This includes managing micro-batches for each module separately and scheduling computation and memory copies.

\mypar{Search Space}
Given the formulation above, we construct a search space for possible system hyperparameters that affect the execution time $T$ given accumulated batch size $B$ (shown in~\cref{eq:optim-goal}).
The time is a function of components as in~\cref{sec:background}.
The variables in the search space that would influence the throughput are shown in~\cref{tab:Notation-search-space}. 
\begin{align}
   \underset{B, b_{a}, b_{e}, \omega, S_\textrm{expert},S_\textrm{Params}}{\text{maximize}} \quad  \frac{B}{T(B, S_\textrm{Expert},S_\textrm{Params}, b_{a}, b_{e}, \omega)} \label{eq:optim-goal}
\end{align}
This search space is constrained by host memory capacity and GPU memory capacity as in~\cref{eq:batch-size-cpu,eq:batch-size-gpu}
\begin{align}
   S_\textrm{KV-CPU}(B) &+ S_\textrm{Model} \le m_c \label{eq:batch-size-cpu} \\
   S_\textrm{Params} &+ S_\textrm{Expert} + S_\textrm{Dense} \nonumber\\ 
     &+ S_\textrm{KV-GPU}(b_{a}) + S_\textrm{IS}(B,b_{a},b_{e}) \le m_g \label{eq:batch-size-gpu}
\end{align}

\begin{table}[t]
    \centering
    \small
    \begin{tabular}{|c|c|}
    \hline
    Notations        & Meaning \\
    \hline
    \hline
    \multicolumn{2}{|c|}{\textit{System \& Model Parameters (Profiled)}} \\
    \hline
     $m_{c}$               &Host memory capacity     \\
     \hline
     $m_{g}$               &GPU memory capacity    \\
     \hline
     $S_{\text{Model}}$           &Size of the model    \\
     \hline
     \hline
     \multicolumn{2}{|c|}{\textit{Functions}} \\
     \hline 
    $S_{\text{KV-CPU}}$      & Size of KV cache in CPU memory. \\
    \hline
    $S_{\text{KV-GPU}}$      & Size of KV cache in GPU memory. \\
    \hline
    $T$               &Execution time for a batch $B$      \\
    \hline
    $S_{\text{IS}}$          & Size of intermediate states in execution \\
    \hline
    \hline
     \multicolumn{2}{|c|}{\textit{Constants (Predetermined)}} \\
     \hline 

    $S_{\text{Dense}}$        & GPU prefetch buffer size for dense modules\\
     \hline
     \hline
     \multicolumn{2}{|c|}{\textit{Variables}} \\
     \hline 
    $B$               &Accumulated batch size for sparse MoE layer   \\
    \hline
    $b_{a}$           & GPU attention module batch size  \\
    \hline
    $b_{e}$           & GPU expert module batch size \\
    \hline
    $\omega$          & Split ratio of $B$ to CPU for attention module\\
    \hline
    $S_{\text{Expert}}$      & Reserved GPU buffer size for expert modules \\
    \hline    
    $S_{\text{Params}}$          & Size of cached model parameters in the GPU \\
    \hline
    \end{tabular}
    \caption{Notations for the search space.}
    \label{tab:Notation-search-space}
\end{table}

We consider the following factors in the search space:

\textbf{P-D disaggregation.}
A widely adopted approach, known as the prefill-decode (P-D) disaggregation~\cite{splitwise,vllm}, defines two classes of DAGs. During the prefill phase, there is no HtoD KV-cache copy in the DAG, whereas the decoding phase considers all possible nodes, as shown in~\cref{fig:DAG}. The parameter $B$ has minimal effect on $S_{\text{IS}}$ during decoding, since the hidden states it influences is typically sized in MBs, thus incurring negligible overheads. Consequently, we set $B$ in the decoding phase to the maximum value permitted by the host memory size.

 \textbf{Module micro-batch size $b_{a}$ and $b_{e}$.}
Choosing $b_a$ presents a tradeoff between two types of overhead: (i) the traffic on HtoD link due to insufficient cached model weights with larger $b_a$, and (ii) GPU underutilization due to attention kernel launch with smaller $b_a$.
Larger $b_a$ uses more space for intermediate states in the GPU, squeezing the space for expert buffer, KV-cache buffer and cached model parameters.
Smaller $b_a$ reduces arithmetic intensity, risking underutilization of GPU computations as shown in~\cref{sec:background}.

Similar consideration applies to the selection of $b_{e}$. Furthermore, while we estimate the number of tokens routed to each expert by assuming an even distribution, the actual number remains unknown until runtime. Therefore, \(b_{e}\) is chosen to prevent out-of-memory (OOM) errors.

\textbf{Accumulated batch split ratio to CPU $\omega$.}
Choosing the split ratio considers the following factors:
(i) CPU computation latency \textit{Self\_Attn}, comparing to (ii) HtoD copy overhead of the KV-cache together with GPU computation latency and (iii) bandwidth for expert prefetching.
CPU-based computation reduces the HtoD overhead and its bandwidth usage as the KV-cache stays in Host memory.
CPU does not need to finish at the same time or earlier as GPU as in conventional load balanced scheduling, bandwidth saving for prefetching large amount of expert can be better.

 \textbf{Size of reserved GPU buffer $S_{\text{Expert}}$.} We aim to select an appropriate size that allows the HtoD engine to prefetch expert weights into the GPU whenever there are periods of idle PCIe time, without overly consuming scarce GPU memory.

 \textbf{Size of cached model parameters in GPU $S_{\text{Params}}$.} It is possible that the entire process is memory-bound, so adding $S_{\text{Expert}}$ or $b_{a}$ may not yield any benefit. In this case, using the spare GPU space to cache part of the model parameters reduces HtoD traffic for copying model weights, thereby alleviating memory-bound constraints.

\subsection{Searching Batching Strategy}

\begin{figure}[t]
    \centering
    \includegraphics[width=\linewidth]{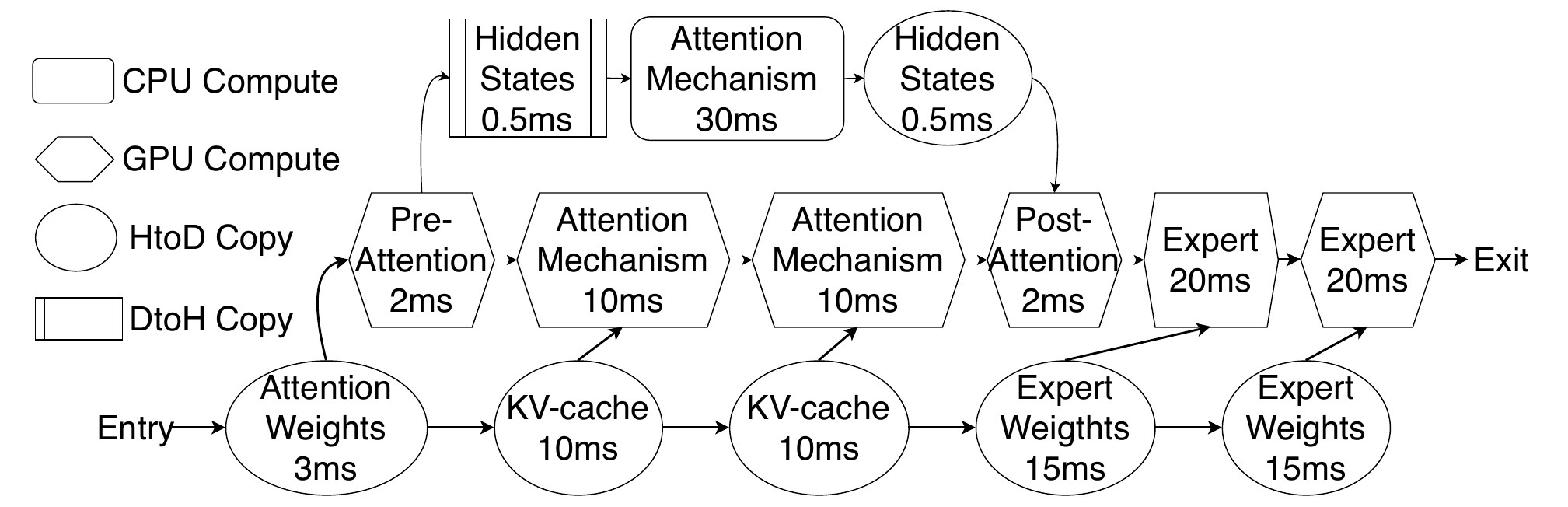}
    \vspace{-0.25in}
    \caption{MoE offloading DAG for module-based batching. The dependencies are denoted with arrows.}
    \label{fig:DAG}
\end{figure}

\mypar{MoE offloading as a DAG}
We can view the model inference as a Directed Acyclic Graph (DAG) of jobs. 
The edges of the DAG represent job dependencies, and each node has the following attributes: i) job execution time, being determined by model, hardware platform and context length, ii) type of the job, being either computation or memory copy.
\cref{fig:DAG} shows such an example, with the assumption that CPU and GPU run the attention module in parallel followed by memory copy and execution of the expert module.

In the decoding phase, the processing starts by copying attention module weights (if offloaded) to GPU memory. Then the GPU computation for the attention mechanism depends on the results from \textit{Pre-Attention} and corresponding \textit{KV-cache} copy. Meanwhile, the CPU kernel for attention mechanism would launch and only depends on Pre-Attention results because the KV-cache are fully offloaded to host memory and CPU kernel can access it directly. After batches of attention mechanism are completed, the results from both CPU and GPU are concatenated and pass through the \textit{Post-Attention} stage. In the followed sparse MoE layer, experts are sequentially executed dependent on the copy of their module weights.
and GPU computation for self-attention mechanism depends on the results of Pre-Attention.

\mypar{Solving minimal runtime on DAG}
With DAG, we abstract the problem of estimating the time consumption of inference to Dynamic Programming for the Longest Path (Critical Path). Let $\text{dp}[v]$ represent the earliest finishing time of node $v$ in the DAG. Initialize $\text{dp}[v]=0$ for all nodes $v$ except the entry node. The $\text{dp}[entry]$ is set to the time cost of itself. Traverse the nodes in topological order, for each node $v$:
\begin{align}
    \text{dp}[v]=\max_{u \in \text{predecessors}(v)} \bigl(\text{dp}[u]\bigr) 
  + \text{cost}(v).
\end{align}
The $\text{dp}[exit]$ represents the finishing time for this DAG.

\section{Evaluation}

\subsection{Experiments Setup}\label{sec:Experiment-Setup}

\mypar{Hardware} 
We conduct benchmarks on various hardware settings as shown in~\cref{tab:testbed-configurations}.
We aim to benchmark the system performance under various memory capacity, GPU and CPU computational power.
C3 has more powerful GPU and larger GPU memory comparing to C1 C2 and C3, while the CPU is weaker.

\begin{table}[t]
    \centering
    \resizebox{\linewidth}{!}{%
    \begin{tabular}{|c|c|c|c|}
    \hline
    Config.    & GPU   & CPU   & Host Memory \\
    \hline
    C1   & A5000 24GB & AMD 7453 28-Core & 256GB \\
    C2   & A5000 24GB & AMD 7453 28-Core & 512GB \\ 
    \hline
    C3    & A6000 48GB & AMD 7313P 16-Core & 480GB \\
    \hline
    \end{tabular}
    }
    \caption{Testbed configurations.}
    \label{tab:testbed-configurations}
\end{table}

\mypar{Datasets}
We used a large variety of LLM tasks that covers long prompting (for prefilling performance) or long output tokens (for decoding performance), including LongBench~\cite{longbench} for long context generation, MMLU~\cite{mmlu} for multiple choice questions, aiming for prefilling and GSM8K~\cite{gsm8k} for math problems and Chatbot-Arena~\cite{chatbot-arena} for multi-round chat.

\mypar{Models} 
We include popular open-sourced MoE models in our evaluations including Mixtral-8x7B, Mixtral-8x22B, DeepSeekV2-236B and DeepSeekR1-671B.
We exclude text-to-text generation models such as NLLB-MoE~\cite{nllb}, as the output length has less variation in response to the input prompt length.
OLMoE~\cite{olmoe} and Phi~\cite{phi-moe} do not require offloading, thus are omitted in the experiment.

\mypar{Baselines}
We evaluate \sys and \sys’s variant for breakdown, comparing them against baseline systems that
support running LLMs with offloading: 
(i)~\textbf{Lllama.cpp} and \textbf{vLLM} which support continuous batching; 
(ii)~\textbf{DeepSpeed}, \textbf{FlexGen*} and \textbf{MoE-Lightning*} which support model-based batching;
FlexGen does not support MoE models and MoE-Lightning has not yet open sourced, so we replicate their performance using our own implementation.
We choose the inference-specific version of DeepSpeed as the baseline.
Mixtral-Offloading only supports the Mixtral model and has inferior performance than DeepSpeed, thus we omit it from evaluation.

The baselines are compared with two versions of \sys: (i) \textbf{\sys}(G) which only uses GPU for all computation, and (ii) \textbf{\sys}(H) which incorporates the CPU attention for expert offloading bandwidth savings.
\sys(G) aligns with most baselines that all computations are performed in GPU, except for Llama.cpp, FlexGen and MoE-Lightning, which align with \sys(H).

All baseline (vLLM, Llama.cpp, MoE-Lightning) and versions of \sys using requests padded to the maximum prompt length and with continuous batching disabled (when possible), as aligned with FlexGen.

\subsection{End-to-end Performance}\label{sec:end-to-end performance}
\begin{table}[t]
    \centering
    % \small
    \resizebox{\linewidth}{!}{%
    \begin{tabular}{|r|c|c|c|}
    \hline
                        &MMLU   & GSM8K     & ChatBotArena \\
$\rightarrow$Num. Sequences   &116K   & 8.5K        &   36K            \\

(Prompt Len, Decoding Len)   &(512,1)   & (512,256)        &   (256,512)           \\

    \hline

    Llama.cpp       &  149hr       & 374hr           &    6423hr     \\
    vLLM            &   112hr      & 303hr          &   5205hr       \\

    \hline
    DeepSpeed       &   23hr  & 115hr    &     1710hr            \\
    FlexGen*        &   25hr      & 122hr               &   5132hr       \\
    MoE-Lightning*  &    {23hr}      &    {68hr}            &     5123hr     \\ 
    \hline
    \textbf{\sys}(G)    &  \underline{18hr}       &    \underline{12hr}    & \underline{124hr}\\
    \textbf{\sys}(H)    &   \textbf{18hr}  &  \textbf{8hr}  & \textbf{82hr}\\
    \hline
    \end{tabular}
    }
    \caption{Time to complete dataset on C2 with Mixtral-8x22B, including model loading time.}
    \label{tab:time-dataset}
\end{table}

\mypar{Time to complete a dataset}
We evaluate the total time for \sys and the baselines to complete offline inference on large datasets. Note that MMLU only has a prefilling phase, where the first tokens are used as the answer. \cref{tab:time-dataset} reports the performance. 
For runtime larger than 24hr, we estimate the overall time using partial dataset.

{Mixtral-8x22B} is relatively dense, given its “top-2 from 8” expert routing policy. Under traditional model-based batching, with an input prompt length of 512 and a small batch size of 16, each expert is routed $2^{11}$ tokens in the prefill phase, reaching the saturation region (see \cref{fig:expert-arithmetic-intensity}). Therefore, on prefilling-only task like MMLU, \sys provides smaller gains(1.3-1.4$\times$) compared to decoding tasks such as the rest three datasets. For a much sparser model like {DeepSeek-V2} (with a “top-6 from 160” expert routing policy), the performance of baselines during prefilling degrades dramatically, as shown in \cref{tab:prefill-performance}.

In the decoding phase, model-based batching yields a low computation workload per expert. Therefore, \sys(H) significantly outperforms the baselines.  On GSM8K, \sys(H) achieves a 9–15$\times$ speedup, and when the decoding length is doubled in ChatbotArena, \sys(H) improves by 21–63$\times$.
The performance difference between \sys(G) and \sys(H) underscores the effectiveness of our engine in leveraging CPU computation.

\mypar{Cost savings}
We demonstrate that \sys enables cost savings by exchanging GPU costs and power with host memory utilization. \sys supports models with hundreds of GB in size and thousands in batch size on a single GPU workstation. Previously, users needed to invest in expensive and power-hungry 8-GPU servers.
\cref{tab:cost-efficient} presents the results. Instead of investing in powerful 8-GPU servers and underutilizing host memory, the same workload can be deployed across eight memory-enhanced servers, each equipped with a single GPU.

As shown in~\cref{tab:cost-efficient}, \sys achieves a comparable throughput of an 8-GPU server with a single GPU, while requiring only 21\% of the infrastructure budget and corresponding power. The host memory's low power consumption is the key to the efficiency (in~\cite{moe-cap}). This configuration also achieves comparable throughput.
% \fixme{Change to deepseek-V2}
\begin{table}[t]
    \centering
    \small
    \begin{tabular}{|c|c|c|c|}
    \hline
        Throughput  & Config. & Power & Cost \\
    \hline
    \hline
    \multirow{4}{*}{\centering 140}
           & 8xNVIDIA-A5000 & 1600W & 20K\$ \\
             & 1xAMD-7453 & 100W & 1.2K\$ \\
             & 512GB Host & 80W & 1.1K\$ \\
    \hline
        vLLM &   & 1780W & 22.3K\$ \\
    \hline
    \hline
    \multirow{4}{*}{\centering 143}
           & 1xNVIDIA-A5000 & 200W & 2.5K\$ \\
             & 1xAMD-7453 & 100W & 1.2K\$ \\
             & 512GB Host & 80W & 1.1K\$ \\
    \hline
        \textbf{\sys}  &  & \textbf{380W} & \textbf{4.8K}\$\\
    \hline
    \end{tabular}
    \caption{Server settings to save cost. Mixtral-8x22B.}
    \label{tab:cost-efficient}
\end{table}

\subsection{\sys Breakdown}

\mypar{Decoding throughput}
We evaluate the total decoding throughput under an increasing KV-cache size throughout the decoding process. \cref{tab:decode-performance} presents the results. Overall, \sys achieves a 16-31$\times$ throughput improvement over the baselines, showing more pronounced gains for models with higher sparsity, such as {DeepSeek}. Note that, for {DeepSeek}, the latent KV-cache must be up-projected by a factor of 71 at runtime. Performing this projection on the CPU or copying the projected results from the device to the host (DtoH) introduces substantial overhead. Consequently, for {DeepSeek-V2}, the split ratio $w$ is set to zero, which aligns with the configuration used by \sys(G).

\begin{table}[t]
\centering
\resizebox{\linewidth}{!}{%
\begin{tabular}{|r|cc|cc|cc|cc|}
\hline
% & \multicolumn{8}{c|}{C2} & \multicolumn{6}{c|}{C3} \\
% \cline{2-13}
$\rightarrow$Models & \multicolumn{2}{c|}{Mixtral 8x7B} & \multicolumn{2}{c|}{Mixtral 8x22B}  & \multicolumn{2}{c|}{DeepSeek-V2 236B} & \multicolumn{2}{c|}{DeepSeek-R1 671B}\\
\cline{1-9}

$\rightarrow$Decode Len  & 256  & 1024  & 256  & 1024  & 256 & 1024  &256 & 1024 \\
% \cline{2-13}
% $\rightarrow$Bsz &  &  &  &  &  &  &  &  &  &  &  & \\
\hline
% % Mixtral-Offloading  &8      &       &       &$\xmark$  &       &       &$\xmark$    &       &     \\ 
Llama.cpp           &4        &   3    &2           &   0.8    & 1   &   0.3   & 0.9& $<$0.1\\
vLLM                &31        &    14   & 2           &    1   &        0.8    & $<$0.1 & Fail & Fail\\ 

 \hline

DeepSpeed           &27          &  26     &4              &   3    &1              &   1   & Fail&Fail\\
FlexGen*            &33         & 30      &5            &   4    &   1              &   1    & Fail&Fail\\ 
MoE-Lightning*      &89           &  78     &9             &  6     &  1      &    1    & Fail  &Fail\\
\hline
 \textbf{\sys}(G)    & \underline{195}       & \underline{93}   & \underline{54}       & \underline{27}     &\underline{31}           &\underline{16}        &\underline{17}    & \underline{9}\\
\textbf{\sys}(H)    &\textbf{469}        &\textbf{283}   &\textbf{91}    &\textbf{57}   &\textbf{31}       &\textbf{16}   &\textbf{17} &\textbf{9} \\
\hline

\end{tabular}
}
\caption{Decoding throughout (token/s) on C2 with prompt length 512.}
\label{tab:decode-performance} 
\end{table}

\mypar{Prefill throughput} \cref{tab:prefill-performance} reports the performance.
\sys not significantly outperform DeepSpeed as in Mixtral type of models with less sparsity. \sys achieves 7.2$\times$ more throughput with DeepSeek with higher sparsity.

Batch size in DeepSpeed is bounded by attention peak memory, limiting the batch size to only 8 while DeepSeek-V2 has 160 experts per layer. Thus each expert only has less than 10 tokens to process, leading to GPU computation capability being wasted. FlexGen and MoE-Lightning have the same problem. They can reuse the copied model weights, thus slightly better than DeepSpeed. 

\begin{table}[t]
    \centering
\resizebox{\linewidth}{!}{%
    \begin{tabular}{|r|c|c|c|c|}
    \hline
$\rightarrow$Model      & Mixtral   & Mixtral   & DeepSeekV2  & DeepSeekR1  \\
$\downarrow$Baselines   & 8x7B      & 8x22B     & 236B        & 671B  \\
    \hline
    Llama.cpp           &   328     &  110      &  23         & 6    \\
     vLLM               &   1347     &  147       &  97         & Fail    \\
     \hline

    DeepSpeed           &   2621    &  710      & 109         & Fail  \\
    FlexGen*            &   2199    &  655      &  77         & Fail     \\
    MoE-Lightning*      &   2237    &  702      &  98         & Fail     \\
    \hline
    \textbf{\sys}    &   \textbf{2790}    & \textbf{907}       & \textbf{787}    &\textbf{204}       \\
    \hline
    \end{tabular}
}
    \caption{Prefill throughput (token/s) on C2 with prompt length 512. \sys outperforms on baselines on larger and more sparse MoE. In prefilling phase \sys(G) and \sys(H) has same performance using only GPU.}
    \label{tab:prefill-performance}
\end{table}

\begin{table}[t]
    \centering
     \resizebox{\linewidth}{!}{%
    \begin{tabular}{|r|cc|cc|cc|cc|}
    \hline
    & \multicolumn{2}{c|}{16K-8K} & \multicolumn{2}{c|}{8K-16K} & \multicolumn{2}{c|}{8K-4K} & \multicolumn{2}{c|}{4K-2K} \\
    % \cline{2-9}
    & \multicolumn{2}{c|}{$B=50$} & \multicolumn{2}{c|}{$B=50$} & \multicolumn{2}{c|}{$B=100$} & \multicolumn{2}{c|}{$B=200$} \\
    \cline{2-9}
    & P         & D      & P        & D       & P        & D       & P        & D \\
    % & TP              & TP            & TP             &TP              &TP              &TP                &TP              &TP\\
    \hline
        % Llama.cpp  &    12\fixme{X}    &    4   &    12   &   4    &    12   &   5    &  14     &  6\\
        vLLM       &    1182    &   1    &  1329    &   1    &   1325    &   1    &    1359   & 1 \\
        \hline
        DeepSpeed  &   2617     &   1    &   2621    &   1    &   2621    &  2     & 2653      & 3\\
        FlexGen*   &   2173     &    2   &   2187    &   2    &   2187    &   3    &    2192   & 5\\
          MoE-Lightning*   &  2218      &    2   &   2221    &    2   &  2221     &   4    &   2232    & 6\\
    \hline
        \textbf{\sys(H)} &\textbf{2662}        &\textbf{13}      &\textbf{2684}       &\textbf{13}      &\textbf{2686}       &\textbf{20}      &\textbf{2667}       &\textbf{50}       \\    
    \hline
    \end{tabular}
    }
    \caption{Throughput for long context generation on C1 with Mixtral-8x7B using Longbench. 16K-8K represents prefill (P) length 16K tokens and decoding (D) length 8K tokens. We conducted preliminary trials with Llama.cpp; however, it proved computationally prohibitive in practice while yielding notably lower performance relative to our other baselines. Consequently, it is omitted from our results. }
    \label{tab:longcontext-performance}
\end{table}

\mypar{Long context performance}
We aim to investigate whether \sys maintains its advantages over baselines under long-context scenarios. As shown in \cref{tab:longcontext-performance}, for four representative long-context generation tasks, \sys outperforms the baselines by 7-13$\times$. Although a longer context constrains the maximum batch size that can be cached in host memory, \sys is designed to maintain a larger intermediate state space for computation, thereby supporting much larger batch sizes than the baselines.

For a prompt length of 512 and a decoding length of 256 for Mixtral-8x7B on C1, the decoding throughput of \sys is 364 tokens/s. We observe that \sys’s decoding throughput decreases compared to short-context scenarios. This downgrade arises because, with limited host memory, the largest batch size diminishes as the context length grows. However, the prefill remains stable since a sufficient number of tokens is provided for the longer context.

\subsection{Further Study}
We conducted an extensive study of how insufficient batch sizes, CPU attention ratio ($\omega$), and memory constraints affect throughput in \sys. Our findings demonstrate that with insufficient batch sizes(\eg 1, 32), \sys performs better or is comparable with baselines on different models; an optimal $\omega$ significantly improve GPU utilization and overall performance. We also demonstrate the influence of the computational power of the CPU on $\omega$. Detailed discussions and additional experiments are provided in the \cref{sec:ablation_study}.

\section{Conclusion}

\sys is the first system to enable high-throughput offline inference for large MoE models on a single GPU. Its core innovation, module-based batching, has been extensively evaluated against strong baselines on popular MoE models. \sys has the potential for broad impact, enabling AI developers to efficiently run offline inference on personal machines and significantly lowering the barrier to adopting large MoE models.

\newpage
\newpage

\section*{Impact Statement}

This paper presents work whose goal is to advance the field of Machine Learning. There are many potential societal consequences of our work, none of which we feel must be specifically highlighted here.

\bibliography{example_paper}

\begin{thebibliography}{44}
\providecommand{\natexlab}[1]{#1}
\providecommand{\url}[1]{\texttt{#1}}
\expandafter\ifx\csname urlstyle\endcsname\relax
  \providecommand{\doi}[1]{doi: #1}\else
  \providecommand{\doi}{doi: \begingroup \urlstyle{rm}\Url}\fi

\bibitem[Abdin et~al.(2024)Abdin, Jacobs, Awan, Aneja, Awadallah, Awadalla, Bach, Bahree, Bakhtiari, Behl, Benhaim, Bilenko, Bjorck, Bubeck, Cai, Mendes, Chen, Chaudhary, Chopra, Giorno, de~Rosa, Dixon, Eldan, Iter, Garg, Goswami, Gunasekar, Haider, Hao, Hewett, Huynh, Javaheripi, Jin, Kauffmann, Karampatziakis, Kim, Khademi, Kurilenko, Lee, Lee, Li, Liang, Liu, Lin, Lin, Madan, Mitra, Modi, Nguyen, Norick, Patra, Perez{-}Becker, Portet, Pryzant, Qin, Radmilac, Rosset, Roy, Ruwase, Saarikivi, Saied, Salim, Santacroce, Shah, Shang, Sharma, Song, Tanaka, Wang, Ward, Wang, Witte, Wyatt, Xu, Xu, Yadav, Yang, Yang, Yu, Zhang, Zhang, Zhang, Zhang, Zhang, Zhang, Zhang, and Zhou]{phi-moe}
Abdin, M.~I., Jacobs, S.~A., Awan, A.~A., Aneja, J., Awadallah, A., Awadalla, H., Bach, N., Bahree, A., Bakhtiari, A., Behl, H.~S., Benhaim, A., Bilenko, M., Bjorck, J., Bubeck, S., Cai, M., Mendes, C. C.~T., Chen, W., Chaudhary, V., Chopra, P., Giorno, A.~D., de~Rosa, G., Dixon, M., Eldan, R., Iter, D., Garg, A., Goswami, A., Gunasekar, S., Haider, E., Hao, J., Hewett, R.~J., Huynh, J., Javaheripi, M., Jin, X., Kauffmann, P., Karampatziakis, N., Kim, D., Khademi, M., Kurilenko, L., Lee, J.~R., Lee, Y.~T., Li, Y., Liang, C., Liu, W., Lin, E., Lin, Z., Madan, P., Mitra, A., Modi, H., Nguyen, A., Norick, B., Patra, B., Perez{-}Becker, D., Portet, T., Pryzant, R., Qin, H., Radmilac, M., Rosset, C., Roy, S., Ruwase, O., Saarikivi, O., Saied, A., Salim, A., Santacroce, M., Shah, S., Shang, N., Sharma, H., Song, X., Tanaka, M., Wang, X., Ward, R., Wang, G., Witte, P., Wyatt, M., Xu, C., Xu, J., Yadav, S., Yang, F., Yang, Z., Yu, D., Zhang, C., Zhang, C., Zhang, J., Zhang, L.~L., Zhang, Y., Zhang, Y., Zhang, Y., and
  Zhou, X.
\newblock Phi-3 technical report: {A} highly capable language model locally on your phone, 2024.

\bibitem[Aminabadi et~al.(2022)Aminabadi, Rajbhandari, Awan, Li, Li, Zheng, Ruwase, Smith, Zhang, Rasley, and He]{ds-infer}
Aminabadi, R.~Y., Rajbhandari, S., Awan, A.~A., Li, C., Li, D., Zheng, E., Ruwase, O., Smith, S., Zhang, M., Rasley, J., and He, Y.
\newblock {DeepSpeed}-{Inference}: Enabling efficient inference of transformer models at unprecedented scale.
\newblock In \emph{{SC}}, pp.\  46:1--46:15. {IEEE}, 2022.

\bibitem[Asai et~al.(2023)Asai, Min, Zhong, and Chen]{DBLP:conf/acm/AsaiMZC23}
Asai, A., Min, S., Zhong, Z., and Chen, D.
\newblock Retrieval-based language models and applications.
\newblock In \emph{{ACL} (tutorial)}, pp.\  41--46. Association for Computational Linguistics, 2023.

\bibitem[Bai et~al.(2024)Bai, Lv, Zhang, Lyu, Tang, Huang, Du, Liu, Zeng, Hou, Dong, Tang, and Li]{longbench}
Bai, Y., Lv, X., Zhang, J., Lyu, H., Tang, J., Huang, Z., Du, Z., Liu, X., Zeng, A., Hou, L., Dong, Y., Tang, J., and Li, J.
\newblock {LongBench}: {A} bilingual, multitask benchmark for long context understanding.
\newblock In \emph{{ACL} {(1)}}, pp.\  3119--3137. Association for Computational Linguistics, 2024.

\bibitem[Cao et~al.(2024)Cao, Liu, Griggs, Schafhalter, Liu, Sheng, Gonzalez, Zaharia, and Stoica]{moe-lightning}
Cao, S., Liu, S., Griggs, T., Schafhalter, P., Liu, X., Sheng, Y., Gonzalez, J.~E., Zaharia, M., and Stoica, I.
\newblock {MoE-Lightning}: High-throughput moe inference on memory-constrained gpus, 2024.

\bibitem[Chiang et~al.(2024)Chiang, Zheng, Sheng, Angelopoulos, Li, Li, Zhu, Zhang, Jordan, Gonzalez, and Stoica]{chatbot-arena}
Chiang, W., Zheng, L., Sheng, Y., Angelopoulos, A.~N., Li, T., Li, D., Zhu, B., Zhang, H., Jordan, M.~I., Gonzalez, J.~E., and Stoica, I.
\newblock Chatbot arena: An open platform for evaluating llms by human preference.
\newblock In \emph{{ICML}}. OpenReview.net, 2024.

\bibitem[Cobbe et~al.(2021)Cobbe, Kosaraju, Bavarian, Chen, Jun, Kaiser, Plappert, Tworek, Hilton, Nakano, Hesse, and Schulman]{gsm8k}
Cobbe, K., Kosaraju, V., Bavarian, M., Chen, M., Jun, H., Kaiser, L., Plappert, M., Tworek, J., Hilton, J., Nakano, R., Hesse, C., and Schulman, J.
\newblock Training verifiers to solve math word problems, 2021.

\bibitem[Costa{-}juss{\`{a}} et~al.(2022)Costa{-}juss{\`{a}}, Cross, {\c{C}}elebi, Elbayad, Heafield, Heffernan, Kalbassi, Lam, Licht, Maillard, Sun, Wang, Wenzek, Youngblood, Akula, Barrault, Gonzalez, Hansanti, Hoffman, Jarrett, Sadagopan, Rowe, Spruit, Tran, Andrews, Ayan, Bhosale, Edunov, Fan, Gao, Goswami, Guzm{\'{a}}n, Koehn, Mourachko, Ropers, Saleem, Schwenk, and Wang]{nllb}
Costa{-}juss{\`{a}}, M.~R., Cross, J., {\c{C}}elebi, O., Elbayad, M., Heafield, K., Heffernan, K., Kalbassi, E., Lam, J., Licht, D., Maillard, J., Sun, A., Wang, S., Wenzek, G., Youngblood, A., Akula, B., Barrault, L., Gonzalez, G.~M., Hansanti, P., Hoffman, J., Jarrett, S., Sadagopan, K.~R., Rowe, D., Spruit, S., Tran, C., Andrews, P., Ayan, N.~F., Bhosale, S., Edunov, S., Fan, A., Gao, C., Goswami, V., Guzm{\'{a}}n, F., Koehn, P., Mourachko, A., Ropers, C., Saleem, S., Schwenk, H., and Wang, J.
\newblock No language left behind: Scaling human-centered machine translation, 2022.

\bibitem[Cui et~al.(2023)Cui, Han, Ouyang, Wang, Zheng, Ma, Yang, Yang, Xue, Qiu, Zhou, Chen, Tan, and Guo]{brainstorm}
Cui, W., Han, Z., Ouyang, L., Wang, Y., Zheng, N., Ma, L., Yang, Y., Yang, F., Xue, J., Qiu, L., Zhou, L., Chen, Q., Tan, H., and Guo, M.
\newblock Optimizing dynamic neural networks with brainstorm.
\newblock In \emph{{OSDI}}, pp.\  797--815. {USENIX} Association, 2023.

\bibitem[Dao(2024)]{flashattn-2}
Dao, T.
\newblock {FlashAttention}-2: Faster attention with better parallelism and work partitioning.
\newblock 2024.

\bibitem[DeepSeek{-}AI et~al.(2024)DeepSeek{-}AI, Liu, Feng, Wang, Wang, Liu, Zhao, Deng, Ruan, Dai, Guo, Yang, Chen, Ji, Li, Lin, Luo, Hao, Chen, Li, Zhang, Xu, Yang, Zhang, Ding, Xin, Gao, Li, Qu, Cai, Liang, Guo, Ni, Li, Chen, Yuan, Qiu, Song, Dong, Gao, Guan, Wang, Zhang, Xu, Xia, Zhao, Zhang, Li, Wang, Zhang, Zhang, Tang, Li, Tian, Huang, Wang, Zhang, Zhu, Chen, Du, Chen, Jin, Ge, Pan, Xu, Chen, Li, Lu, Zhou, Chen, Wu, Ye, Ma, Wang, Zhou, Yu, Zhou, Zheng, Wang, Pei, Yuan, Sun, Xiao, Zeng, An, Liu, Liang, Gao, Zhang, Li, Jin, Wang, Bi, Liu, Wang, Shen, Chen, Chen, Nie, and Sun]{deepseek-v2}
DeepSeek{-}AI, Liu, A., Feng, B., Wang, B., Wang, B., Liu, B., Zhao, C., Deng, C., Ruan, C., Dai, D., Guo, D., Yang, D., Chen, D., Ji, D., Li, E., Lin, F., Luo, F., Hao, G., Chen, G., Li, G., Zhang, H., Xu, H., Yang, H., Zhang, H., Ding, H., Xin, H., Gao, H., Li, H., Qu, H., Cai, J.~L., Liang, J., Guo, J., Ni, J., Li, J., Chen, J., Yuan, J., Qiu, J., Song, J., Dong, K., Gao, K., Guan, K., Wang, L., Zhang, L., Xu, L., Xia, L., Zhao, L., Zhang, L., Li, M., Wang, M., Zhang, M., Zhang, M., Tang, M., Li, M., Tian, N., Huang, P., Wang, P., Zhang, P., Zhu, Q., Chen, Q., Du, Q., Chen, R.~J., Jin, R.~L., Ge, R., Pan, R., Xu, R., Chen, R., Li, S.~S., Lu, S., Zhou, S., Chen, S., Wu, S., Ye, S., Ma, S., Wang, S., Zhou, S., Yu, S., Zhou, S., Zheng, S., Wang, T., Pei, T., Yuan, T., Sun, T., Xiao, W.~L., Zeng, W., An, W., Liu, W., Liang, W., Gao, W., Zhang, W., Li, X.~Q., Jin, X., Wang, X., Bi, X., Liu, X., Wang, X., Shen, X., Chen, X., Chen, X., Nie, X., and Sun, X.
\newblock Deepseek-v2: {A} strong, economical, and efficient mixture-of-experts language model.
\newblock \emph{CoRR}, abs/2405.04434, 2024.

\bibitem[Eliseev \& Mazur(2023)Eliseev and Mazur]{mixtral-offload}
Eliseev, A. and Mazur, D.
\newblock Fast inference of mixture-of-experts language models with offloading, 2023.

\bibitem[Fu et~al.(2024)Fu, Jiang, Huang, Nie, Lu, Xue, He, Sit, Xue, Dong, Miao, Zou, Ponti, and Mai]{moe-cap}
Fu, Y., Jiang, Y., Huang, Y., Nie, P., Lu, Z., Xue, L., He, C., Sit, M.-K., Xue, J., Dong, L., Miao, Z., Zou, K., Ponti, E., and Mai, L.
\newblock {MoE-CAP}: Cost-accuracy-performance benchmarking for mixture-of-experts systems, 2024.

\bibitem[Harma et~al.(2024)Harma, Chakraborty, Kostenok, Mishin, Ha, Falsafi, Jaggi, Liu, Oh, Subramanian, and Yazdanbakhsh]{sparse-quant}
Harma, S.~B., Chakraborty, A., Kostenok, E., Mishin, D., Ha, D., Falsafi, B., Jaggi, M., Liu, M., Oh, Y., Subramanian, S., and Yazdanbakhsh, A.
\newblock Effective interplay between sparsity and quantization: From theory to practice, 2024.

\bibitem[He et~al.(2024)He, Zhang, Wang, Yin, Zeng, Shi, Tang, Chu, Tsang, and Ong]{expert-flow}
He, X., Zhang, S., Wang, Y., Yin, H., Zeng, Z., Shi, S., Tang, Z., Chu, X., Tsang, I.~W., and Ong, Y.
\newblock {ExpertFlow}: Optimized expert activation and token allocation for efficient mixture-of-experts inference, 2024.

\bibitem[Hendrycks et~al.(2021)Hendrycks, Burns, Basart, Zou, Mazeika, Song, and Steinhardt]{mmlu}
Hendrycks, D., Burns, C., Basart, S., Zou, A., Mazeika, M., Song, D., and Steinhardt, J.
\newblock Measuring massive multitask language understanding.
\newblock In \emph{ICLR}. OpenReview.net, 2021.

\bibitem[Hwang et~al.(2024)Hwang, Wei, Cao, Hwang, Tang, Cao, and Yang]{pregate-moe}
Hwang, R., Wei, J., Cao, S., Hwang, C., Tang, X., Cao, T., and Yang, M.
\newblock Pre-gated {MoE}: An algorithm-system co-design for fast and scalable mixture-of-expert inference.
\newblock In \emph{{ISCA}}, pp.\  1018--1031. {IEEE}, 2024.

\bibitem[Jiang et~al.(2024{\natexlab{a}})Jiang, Sablayrolles, Mensch, Bamford, Chaplot, de~Las~Casas, Bressand, Lengyel, Lample, Saulnier, Lavaud, Lachaux, Stock, Scao, Lavril, Wang, Lacroix, and Sayed]{mixtral}
Jiang, A.~Q., Sablayrolles, A., Mensch, A., Bamford, C., Chaplot, D.~S., de~Las~Casas, D., Bressand, F., Lengyel, G., Lample, G., Saulnier, L., Lavaud, L.~R., Lachaux, M., Stock, P., Scao, T.~L., Lavril, T., Wang, T., Lacroix, T., and Sayed, W.~E.
\newblock Mixtral of experts, 2024{\natexlab{a}}.

\bibitem[Jiang et~al.(2024{\natexlab{b}})Jiang, Zhou, Cao, Stoica, and Yu]{neo}
Jiang, X., Zhou, Y., Cao, S., Stoica, I., and Yu, M.
\newblock {NEO:} saving {GPU} memory crisis with {CPU} offloading for online {LLM} inference, 2024{\natexlab{b}}.

\bibitem[Kamahori et~al.(2024)Kamahori, Gu, Zhu, and Kasikci]{fiddler}
Kamahori, K., Gu, Y., Zhu, K., and Kasikci, B.
\newblock Fiddler: {CPU-GPU} orchestration for fast inference of mixture-of-experts models, 2024.

\bibitem[Kwon et~al.(2023)Kwon, Li, Zhuang, Sheng, Zheng, Yu, Gonzalez, Zhang, and Stoica]{vllm}
Kwon, W., Li, Z., Zhuang, S., Sheng, Y., Zheng, L., Yu, C.~H., Gonzalez, J., Zhang, H., and Stoica, I.
\newblock Efficient memory management for large language model serving with pagedattention.
\newblock In \emph{{SOSP}}, pp.\  611--626. {ACM}, 2023.

\bibitem[Li et~al.(2023)Li, Jiang, Zhu, Wang, and Xu]{lina}
Li, J., Jiang, Y., Zhu, Y., Wang, C., and Xu, H.
\newblock Accelerating distributed {MoE} training and inference with {Lina}.
\newblock In \emph{{USENIX} Annual Technical Conference}, pp.\  945--959. {USENIX} Association, 2023.

\bibitem[Liu et~al.(2023)Liu, Wang, and Jiang]{janus}
Liu, J., Wang, J.~H., and Jiang, Y.
\newblock Janus: {A} unified distributed training framework for sparse {Mixture}-of-{Experts} models.
\newblock In \emph{{SIGCOMM}}, pp.\  486--498. {ACM}, 2023.

\bibitem[Luan et~al.(2024)Luan, Mao, Wang, Lin, Kamsetty, Chen, Su, Veeramani, Lee, Cho, Zinzow, Liang, Stoica, and Wang]{stream-batch}
Luan, F.~S., Mao, Z., Wang, R.~Y., Lin, C., Kamsetty, A., Chen, H., Su, C., Veeramani, B., Lee, S., Cho, S., Zinzow, C., Liang, E., Stoica, I., and Wang, S.
\newblock The streaming batch model for efficient and fault-tolerant heterogeneous execution, 2024.

\bibitem[Mischler et~al.(2024)Mischler, Li, Bickel, Mehta, and Mesgarani]{mischler2024contextual}
Mischler, G., Li, Y.~A., Bickel, S., Mehta, A.~D., and Mesgarani, N.
\newblock Contextual feature extraction hierarchies converge in large language models and the brain.
\newblock \emph{Nature Machine Intelligence}, pp.\  1--11, 2024.

\bibitem[Muennighoff et~al.(2024)Muennighoff, Soldaini, Groeneveld, Lo, Morrison, Min, Shi, Walsh, Tafjord, Lambert, Gu, Arora, Bhagia, Schwenk, Wadden, Wettig, Hui, Dettmers, Kiela, Farhadi, Smith, Koh, Singh, and Hajishirzi]{olmoe}
Muennighoff, N., Soldaini, L., Groeneveld, D., Lo, K., Morrison, J., Min, S., Shi, W., Walsh, P., Tafjord, O., Lambert, N., Gu, Y., Arora, S., Bhagia, A., Schwenk, D., Wadden, D., Wettig, A., Hui, B., Dettmers, T., Kiela, D., Farhadi, A., Smith, N.~A., Koh, P.~W., Singh, A., and Hajishirzi, H.
\newblock {OLMoE}: Open mixture-of-experts language models, 2024.

\bibitem[Narayan et~al.(2022)Narayan, Chami, Orr, and R{\'{e}}]{DBLP:journals/pvldb/NarayanCOR22}
Narayan, A., Chami, I., Orr, L.~J., and R{\'{e}}, C.
\newblock Can foundation models wrangle your data?
\newblock \emph{Proc. {VLDB} Endow.}, 16\penalty0 (4):\penalty0 738--746, 2022.

\bibitem[{NVIDIA}(2024)]{trt-llm}
{NVIDIA}.
\newblock {TensorRT-LLM}.
\newblock \url{https://github.com/NVIDIA/TensorRT-LLM}, 2024.
\newblock Accessed: 2024-05-17.

\bibitem[{NVIDIA}(2025)]{gemm-nvidia}
{NVIDIA}.
\newblock Matrix multiplication background user's guide.
\newblock \url{https://docs.nvidia.com/deeplearning/performance/dl-performance-matrix-multiplication/index.html}, 2025.
\newblock Accessed: 2025-01-27.

\bibitem[Ollama(2025)]{ollama}
Ollama.
\newblock Ollama.
\newblock \url{https://github.com/ollama/ollama}, 2025.

\bibitem[Patel et~al.(2024)Patel, Choukse, Zhang, Shah, Goiri, Maleki, and Bianchini]{splitwise}
Patel, P., Choukse, E., Zhang, C., Shah, A., Goiri, {\'{I}}., Maleki, S., and Bianchini, R.
\newblock Splitwise: Efficient generative {LLM} inference using phase splitting.
\newblock In \emph{{ISCA}}, pp.\  118--132. {IEEE}, 2024.

\bibitem[{Qwen Team}(2024)]{qwen-moe}
{Qwen Team}.
\newblock Qwen1.5-moe: Matching 7b model performance with 1/3 activated parameters", 2024.
\newblock URL \url{https://qwenlm.github.io/blog/qwen-moe/}.
\newblock Accessed: 2024-05-17.

\bibitem[Sheng et~al.(2023)Sheng, Zheng, Yuan, Li, Ryabinin, Chen, Liang, R{\'{e}}, Stoica, and Zhang]{flexgen}
Sheng, Y., Zheng, L., Yuan, B., Li, Z., Ryabinin, M., Chen, B., Liang, P., R{\'{e}}, C., Stoica, I., and Zhang, C.
\newblock {FlexGen}: High-throughput generative inference of large language models with a single {GPU}.
\newblock In \emph{{ICML}}, volume 202 of \emph{Proceedings of Machine Learning Research}, pp.\  31094--31116. {PMLR}, 2023.

\bibitem[Song et~al.(2024)Song, Zhong, and Chen]{pro-moe}
Song, X., Zhong, Z., and Chen, R.
\newblock {ProMoE}: Fast moe-based {LLM} serving using proactive caching, 2024.

\bibitem[Song et~al.(2023)Song, Mi, Xie, and Chen]{powerinfer}
Song, Y., Mi, Z., Xie, H., and Chen, H.
\newblock {PowerInfer}: Fast large language model serving with a consumer-grade {GPU}, 2023.

\bibitem[Tang1 et~al.(2024)Tang1, Liu, Hou, Pu, Wang, Heng, Li, and Guo]{hobbit}
Tang1, P., Liu, J., Hou, X., Pu, Y., Wang, J., Heng, P.-A., Li, C., and Guo, M.
\newblock Hobbit: A mixed precision expert offloading system for fast {MoE} inference, 2024.

\bibitem[van Renen et~al.(2024)van Renen, Stoian, and Kipf]{DBLP:journals/pvldb/RenenSK24}
van Renen, A., Stoian, M., and Kipf, A.
\newblock Dataloom: Simplifying data loading with llms.
\newblock \emph{Proc. {VLDB} Endow.}, 17\penalty0 (12):\penalty0 4449--4452, 2024.

\bibitem[Xue et~al.(2024)Xue, Fu, Lu, Mai, and Marina]{moe-infinity}
Xue, L., Fu, Y., Lu, Z., Mai, L., and Marina, M.~K.
\newblock {MoE-Infinity}: Activation-aware expert offloading for efficient moe serving, 2024.

\bibitem[Yi et~al.(2023)Yi, Guo, Wei, Zhou, Wang, and Xu]{edge-moe}
Yi, R., Guo, L., Wei, S., Zhou, A., Wang, S., and Xu, M.
\newblock {EdgeMoE}: Fast on-device inference of moe-based large language models, 2023.

\bibitem[Yu et~al.(2022)Yu, Jeong, Kim, Kim, and Chun]{orca}
Yu, G., Jeong, J.~S., Kim, G., Kim, S., and Chun, B.
\newblock Orca: {A} distributed serving system for transformer-based generative models.
\newblock In \emph{{OSDI}}, pp.\  521--538. {USENIX} Association, 2022.

\bibitem[Zhai et~al.(2023)Zhai, He, Ma, Zong, Zhang, and Zhai]{smart-moe}
Zhai, M., He, J., Ma, Z., Zong, Z., Zhang, R., and Zhai, J.
\newblock {SmartMoE}: Efficiently training sparsely-activated models through combining offline and online parallelization.
\newblock In \emph{{USENIX} Annual Technical Conference}, pp.\  961--975. {USENIX} Association, 2023.

\bibitem[Zhao et~al.(2024)Zhao, Yang, Zhu, Zheng, Kasikci, Zhou, Xing, and Stoica]{blend-serve}
Zhao, Y., Yang, S., Zhu, K., Zheng, L., Kasikci, B., Zhou, Y., Xing, J., and Stoica, I.
\newblock {BlendServe}: Optimizing offline inference for auto-regressive large models with resource-aware batching, 2024.

\bibitem[Zheng et~al.(2022)Zheng, Li, Zhang, Zhuang, Chen, Huang, Wang, Xu, Zhuo, Xing, Gonzalez, and Stoica]{alpa}
Zheng, L., Li, Z., Zhang, H., Zhuang, Y., Chen, Z., Huang, Y., Wang, Y., Xu, Y., Zhuo, D., Xing, E.~P., Gonzalez, J.~E., and Stoica, I.
\newblock Alpa: Automating inter- and intra-operator parallelism for distributed deep learning.
\newblock In \emph{{OSDI}}, pp.\  559--578. {USENIX} Association, 2022.

\bibitem[Zhong et~al.(2024)Zhong, Liang, Wang, Wang, Huang, and Li]{adap-moe}
Zhong, S., Liang, L., Wang, Y., Wang, R., Huang, R., and Li, M.
\newblock {AdapMoE}: Adaptive sensitivity-based expert gating and management for efficient moe inference, 2024.

\end{thebibliography}
\bibliographystyle{icml2025}

%%%%%%%%%%%%%%%%%%%%%%%%%%%%%%%%%%%%%%%%%%%%%%%%%%%%%%%%%%%%%%%%%%%%%%%%%%%%%%%
%%%%%%%%%%%%%%%%%%%%%%%%%%%%%%%%%%%%%%%%%%%%%%%%%%%%%%%%%%%%%%%%%%%%%%%%%%%%%%%
% APPENDIX
%%%%%%%%%%%%%%%%%%%%%%%%%%%%%%%%%%%%%%%%%%%%%%%%%%%%%%%%%%%%%%%%%%%%%%%%%%%%%%%
%%%%%%%%%%%%%%%%%%%%%%%%%%%%%%%%%%%%%%%%%%%%%%%%%%%%%%%%%%%%%%%%%%%%%%%%%%%%%%%
\newpage
\appendix

% \onecolumn
\section{Further Details for Evaluation}
\subsection{\sys Further Study}\label{sec:ablation_study}
\mypar{Impact of insufficient batch sizes}
The performance gains of \sys primarily stem from high GPU utilization and enhanced computation-memory overlapping, achievable through large batch sizes. To enable large input batches without running into memory limitations (OOM), we adopt fine-grained memory management techniques alongside design decisions such as full KV-cache offloading, appropriate module buffer sizing, and optimal module micro-batch selection.

These performance advantages diminish as the batch size decreases, which often occurs due to limited host memory capacity. In the extreme scenario of a batch size of one, only a small subset of experts is activated during each forward pass. Since \sys does not employ heuristics to predict expert activations, it defaults to on-demand fetching after the router stage. Despite this limitation, as shown in \cref{tab:insufficient-bsz}, \sys still outperforms baselines for models such as {Mixtral-8x7B} and {DeepSeek-V2-Lite} at batch size one. This observation indicates that existing baselines deliver lower throughput compared to a strategy relying purely on on-demand copying, partly because they are not specifically optimized for latency-sensitive MoE inference and offloading scenarios.

However, for a batch size of 32, \sys does not exhibit performance advantages on {DeepSeek-V2-Lite}. Since the model is relatively small (~30GB), a significant portion of the model and its KV-cache can remain persistently in GPU memory under the {C1} scenario. Many baselines effectively exploit this capability, whereas \sys deliberately offloads more data to host memory to free GPU resources, anticipating larger batch sizes. This design choice introduces additional overhead when working with relatively small models and workloads, thus reducing \sys's comparative advantage in this specific case.

\begin{table}[t]
\centering
\small
\begin{tabular}{|r|cc|cc|}
\hline
$\rightarrow$Models &\multicolumn{2}{c|}{DeepSeek-V2-Lite} & \multicolumn{2}{c|}{Mixtral-8x7B}   \\
\cline{2-5}

$\rightarrow$Batch Size & 1 & 32 & 1 & 32 \\
\hline
vLLM        & 2.1      &   28       & 0.5    & 5   \\
Llama.cpp   & 0.4      &   30       &0.2     &1.1   \\ 
\hline
                           DeepSpeed   & 1.3       &  41       &0.4     &7.7                              \\
FlexGen*    &  0.9         &  35        &  0.3   &  5.2                                   \\ 
MoE-Lightning(p)*  &  1.0     &  37     &     0.4  &  6.1        \\ 
% MoE-Infinity &5.8       &  35       &1.1     &12.6                              \\ 
\hline
% \textbf{\sys}(G)   & 1.0     &8.4       &16.9       &0.36       &2.9       &5.7       &1.1       &9.1       &18.2       \\
\textbf{\sys}(G) &5.0       &35        & 1.0 &33.6  \\
\hline
\end{tabular}
% }
\caption{Decoding throughput for small batch sizes. Prompt length 512. Decoding length 32.}
\label{tab:insufficient-bsz} 
\end{table}

\begin{figure}[t]
    \centering
    \includegraphics[width=0.8\linewidth]{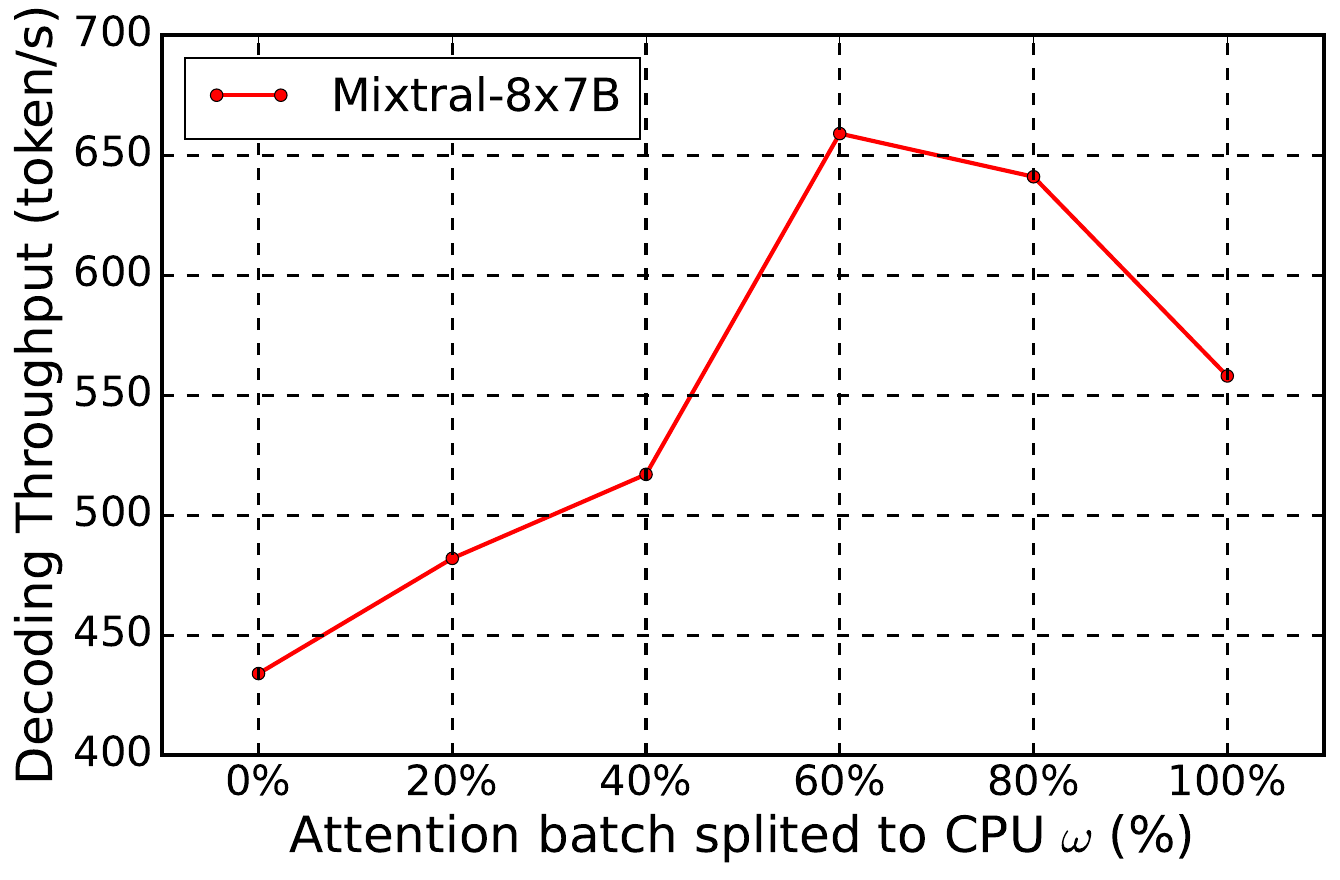}
    \caption{Decoding throughput vs $w$. Prompt length 256, decoding length 32. Mixtral 8x7B. C1. Accumulated batch size $B$ 3640.}
    \label{fig:omega}
\end{figure}

\mypar{Impact of CPU attention ratio $\omega$} The proportion of attention computations offloaded from GPU to CPU ($\omega$) directly influences both the volume of KV-cache data that must be copied to GPU memory and the overall execution time of the CPU-GPU hybrid attention mechanism. We illustrate this relationship for Mixtral-8x7B under scenario C2 in \cref{fig:omega}. In this scenario, pure GPU-based inference is memory-bound. Therefore, offloading part of the attention computation to the CPU reduces the amount of KV-cache data copied, alleviating the memory bottleneck. If the CPU kernel’s execution time for attention is shorter than the combined time required to copy the KV-cache to the GPU and process it there, decoding throughput improves because memory-copy overhead is significantly reduced without making the CPU a bottleneck.

Notably, even when the CPU kernel’s runtime slightly exceeds the GPU's on-demand copy execution time, there can still be throughput benefits, given that memory bandwidth remains the primary limiting factor. However, this advantage holds only up to a certain breakeven point—approximately 60\% CPU offloading in our experiments. Beyond this threshold, memory copying ceases to be the critical bottleneck, and further offloading results in GPU idling as it waits for CPU computations, ultimately leading to degraded overall performance.

\mypar{Influence of CPU computation power} 
In our offloading setting, when the attention module is memory-bound, utilizing the CPU for attention can improve throughput by (1) reducing stress on HtoD memory bandwidth and (2) leveraging CPU computational resources. The efficiency of the CPU kernel for the attention mechanism ultimately determines whether such improvements are realized. Critical factors include vector width, clock frequency, and the number of CPU cores. If a weaker CPU is used, the time spent on CPU-based attention could exceed that of running attention on the GPU—even after copying the KV-cache—resulting in reduced decoding throughput. In such cases, \sys would select $\omega=0$ after the search procedure.

We summarize the attention mechanism workload split ratio under a prompt length of 512 and a decoding length of 256 on Mixtral-8x7B in \cref{tab:omega}, where we simplify the search space for $\omega$ to values from 1/10 to 10/10. For C3, the CPU has fewer cores than C1 and C2 limiting its capacity to handle the same workload without becoming a performance bottleneck. 

\begin{table}[t]
    \centering
    \begin{tabular}{|c|c|c|c|c|}
    \hline
      CPU:GPU   & C1 & C2 & C3 \\
    \hline
      Mixtral-8x7B  &6:4 & 6:4  & 3:7\\
      Mixtral-8x22B & N/A  & 7:3   & 2:8\\
      DeepSeekV2-236B & N/A  &0:10     & 0:10\\
    \hline
    \end{tabular}
    \caption{Attention mechanism workload split ratio. CPU:GPU. Prompt length 512, decoding length 256 on Mixtral-8x7B. C1 cannot hold the model size of Mixtral-8x22B and DeepSeek-V2.}
    \label{tab:omega}
\end{table}
\section{Further Details for Implementation}
\label{sec:appendix-implementation}
\sys is a throughput-oriented offline inference system designed for single-GPU deployments. Its backend consists of approximately 3,000 lines of C++ and 2,000 lines of Python code. We implement a CPU-based attention kernel utilizing AVX intrinsics. Additionally, \sys integrates seamlessly with the HuggingFace generation pipeline, currently supporting the greedy decoding strategy

\mypar{Numerical Consistency of CPU Attention} We implement the widely adopted Grouped Query Attention using the BF16 format with Advanced Vector Extensions (AVX). However, native hardware support for BF16 is limited to relatively recent, high-end CPU generations. To ensure numerical consistency with PyTorch's GPU attention, we represent BF16 data using FP32, explicitly setting all trailing mantissa bits to zero. All computations and accumulations are performed in FP32 precision. After each dot-product accumulation, the results are rounded according to BF16 rounding rules, and the trailing mantissa bits are reset to zero. This approach ensures numerical consistency comparable to GPU-based attention implementations.

\mypar{Workload profiling}
To effectively support DAG-based scheduling optimization, precise workload profiling is required, specifically, the computational latency and peak memory usage for each module, including the self-attention and expert modules. Latency measurements can be obtained by instrumenting each module's forward pass with CUDA events, while peak memory usage—including CUDA context, KV cache, and activations—is measured using the torch memory stats related APIs. Additionally, PCIe bandwidth can be profiled throughput using timed cudaMemcpy operations. Before runtime, each module is profiled offline across various batch sizes and sequence lengths, generating comprehensive profiling data.

%%%%%%%%%%%%%%%%%%%%%%%%%%%%%%%%%%%%%%%%%%%%%%%%%%%%%%%%%%%%%%%%%%%%%%%%%%%%%%%
%%%%%%%%%%%%%%%%%%%%%%%%%%%%%%%%%%%%%%%%%%%%%%%%%%%%%%%%%%%%%%%%%%%%%%%%%%%%%%%
\end{document}